\documentclass[pra,twocolumn,superscriptaddress,notitlepage]{revtex4-1}
\usepackage{graphicx,amsmath,amsfonts,amssymb,upgreek,txfonts,color}
\usepackage[colorlinks,linkcolor=blue,citecolor=blue,urlcolor=blue,breaklinks=true]{hyperref}

\begin{document}

\title{Effects of quadratic coupling and squeezed vacuum injection in an optomechanical cavity assisted with a Bose-Einstein condensate }

\author{A. Dalafi }
\email{a\_dalafi@sbu.ac.ir}
\address{Laser and Plasma Research Institute, Shahid Beheshti University, Tehran 19839-69411, Iran}

\author{M. H. Naderi}
\email{mhnaderi@phys.ui.ac.ir}
\address{Quantum Optics Group, Department of Physics, Faculty of Science, University of Isfahan, Hezar Jerib, 81746-73441, Isfahan, Iran}

\author{Ali Motazedifard}
\email{motazedifard.ali@gmail.com}
\address{Quantum Optics Group, Department of Physics, Faculty of Science, University of Isfahan, Hezar Jerib, 81746-73441, Isfahan, Iran}

\date{\today}

\begin{abstract}
We investigate theoretically a hybrid system consisting of a Bose-Einstein condensate (BEC) trapped inside a laser driven membrane-in-the-middle optomechanical cavity assisted with squeezed vacuum injection whose moving membrane interacts both linearly and quadratically with the radiation pressure of the cavity. It is shown that such a hybrid system is very suitable for generating strong quadrature squeezing in the mechanical mode of the membrane and  the Bogoliubov mode of the BEC in the unresolved sideband regime.  More interestingly, by choosing a suitable sign for the quadratic optomechanical coupling (QOC) one can achieve a very high degree of squeezing in the mechanical mode and a strong entanglement between the mechanical and atomic modes without the necessity of using squeezed light injection. Furthermore, the QOC changes the effective oscillation frequencies of both the mechanical  and the atomic modes and affects their relaxation times. It can also make the system switch form optical bistability to tristability.
\end{abstract}

\pacs{67.85.Hj, 03.75.Gg, 42.50.Wk, 42.50.Lc} 

\maketitle

\section{Introduction}
Generation of squeezed states using cavity optomechanical systems has been one of the most attractive fields of research in the realm of quantum optics over the recent years \cite{kipva, meys ann, asp mod}. If a quantum oscillator is in a state in which the quantum fluctuations of one of its quadratures is suppressed below zero-point level, that state is called a squeezed one \cite{zubairy}. Achieving such states is very important for a variety of applications in ultrasensitive measurements \cite{cave, abram, lahaye, motazedi}.

There are several well-known methods and techniques to achieve stationary squeezing in the conventional optomechanical systems with linear radiation pressure interaction, including periodic modulation of light fields \cite{mari09, farace, aliDCE1, liao, schmidt} or parametrical driving the mechanical oscillator \cite{wl} and coherent modulation of the input laser power \cite{aliDCEJOSAB} which at best lead to $50\%$ noise reduction  below the zero-point level (the so-called 3 dB limit) \cite{mw}. However, in order to surpass this limit one needs to resort to more complex methods of quantum measurements and feedback processes which are difficult to implement \cite{bragi, rus, cl, sz11, sz13}. In recent years, some simpler and more feasible schemes have been proposed to produce strong stationary squeezed states in the optomechanical cavities using quadratic optomechanical coupling (QOC) between the optical and mechanical modes \cite{nunn, sainadh A}. It has been shown \cite{sainadh M, asjadv} that the 3 dB limit in the mechanical squeezing can be beaten through the effect of QOC.

On the other hand, an alternative path to the investigations of cavity optomechanics has been provided recently by systems consisting of ultracold atomic ensembles trapped in optical cavities \cite{Brenn Nature,Gupta,Maschler2008, dom JOSA}.  Besides, the hybrid systems consisting of Bose-Einstein condensates (BECs) exhibit optomechanical properties \cite{Bha 2009,Bha 2010} where the excitation of a collective mode of the condensate plays effectively the role of the vibrational mode of a moving mirror \cite{Kanamoto 2010,dalafi1, dalafi2}. One of the most interesting features of such hybrid systems containing the BEC, is the nonlinear effect of the atom-atom interaction which may lead to the squeezing of the matter field of the BEC \cite{dalafi3}, changeing the pattern of normal modes of the cavity output optical field \cite{dalafi5}, and the generation of squeezed Casimir photons and Bogoliubov/mechanical-type phonons \cite{DCE-BEC}. It has been also shown experimentaly that QOC can be realized in a hybrid cavity containing ultracold atoms \cite{purdy}.

Motivated by the above-mentioned studies, in this paper we study a hybrid system consisting of a BEC trapped inside a membrane-in-the-middle optomechanical cavity. The cavity is injected by a laser pump and a broadband squeezed vacuum field and the membrane  interacts both linearly and quadratically with the radiation pressure of the cavity. It is shown that the QOC not only modifies the quantum fluctuations but also affects the mean-fields of the system. In fact, it changes the effective oscillation frequencies of both the mechanical mode of the membrane and the Bogoliubov mode of the BEC and also affects their relaxation times. Besides, it can make the system switch form optical bistability to optical tristability.

On the other hand, another well-known method for beating the standard squeezing limit of 3 dB for a mechanical mode is the injection of a squeezed light directly into the cavity \cite{sq1}. In this work, we show that the squeezed vacuum light injection can enhance quadrature squeezing in both the mechanical and Bogoliubov modes. Nevertheless, by using a suitable sign of the QOC one can achieve a considerable degree of squeezing (up to 10 dB) in the mechanical mode even in the absence of the squeezed light injection. Moreover, our motivation to equip the optomechanical cavity with the BEC is based on the fact that such a hybridized system can operate effectively in the unresolved sideband regime while most of the schemes proposed for the generation of strong optomechanical squeezing work in the resolved sideband regime \cite{kron, sainadh M, nunn}.

Furthermore, in hybrid BEC-optomechanical systems establishing a strong stationary entanglement between the mechanical and the atomic modes is very challenging due to their indirect interaction  \cite{Rogers}. Here, we show that although the injection of the squeezed light into the optomechanical cavity can strengthen the BEC-membrane entanglement, for a suitable sign of the QOC the entanglement between the mechanical and atomic modes can be increased to very high extent in the dispersive regime even in the absence of the squeezed vacuum injection.

The paper is structured as follows. In Sec.~\ref{Sec.II} we derive the Hamiltonian of the system. In Sec.~\ref{Sec.III} we describe the system dynamics in the framework of the quantum Langevin equations (QLEs). The dynamical evolution of the mean fields of the system and the optical multistability are studied in Sec.~\ref{Sec.IV} and Sec.~\ref{Sec.V}, respectively. In Sec.~\ref{Sec.VI} we investigate the quantum fluctuations of the system in the steady state. Finally, our conclusions are summarized in Sec.~\ref{Sec.VII}.

\section{System Hamiltonian \label{Sec.II}}
\begin{figure}
	\centering
	\includegraphics[width=3in]{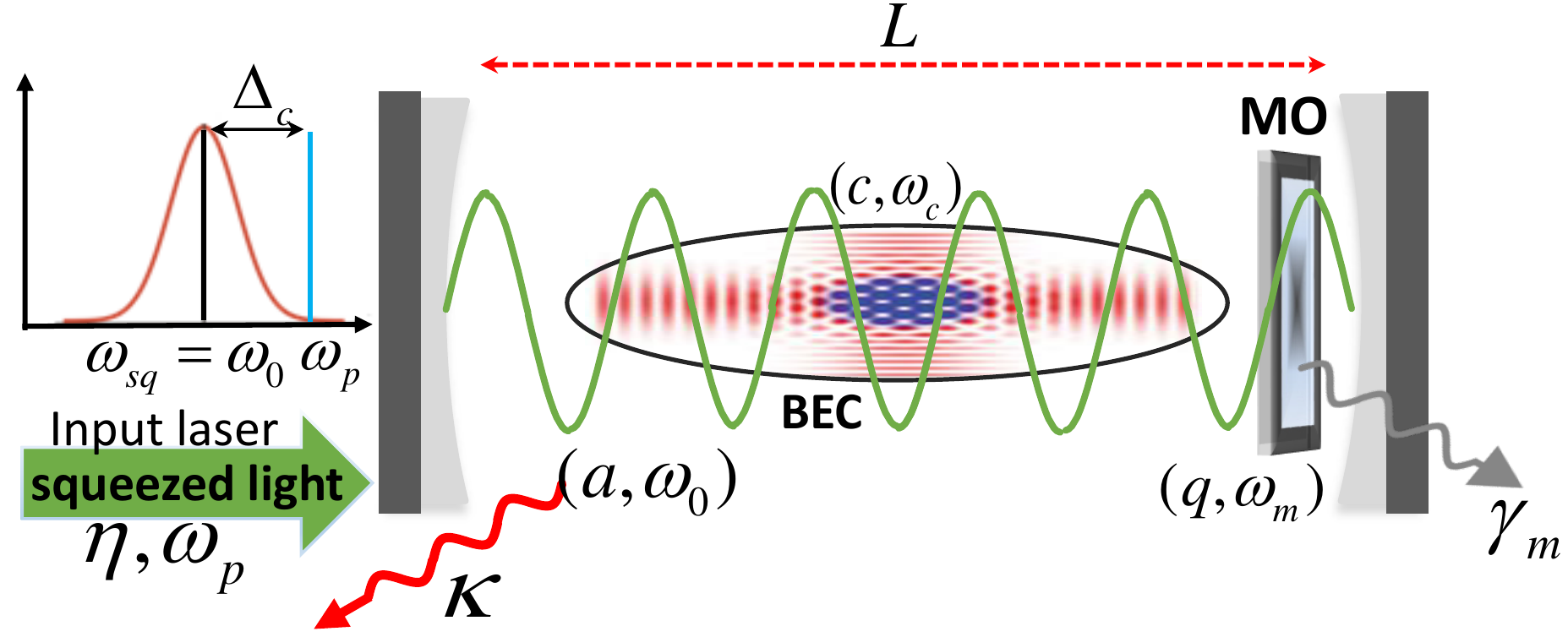} 
	\caption{(Color online) A BEC trapped in a membrane-in-the-middle optomechanical cavity interacting with a single cavity mode. The cavity which decays at rate $\kappa$ is driven through the left mirror by a laser with frequency $\omega_{p}$ and also is injected by a squeezed vacuum field. The membrane is free to oscillate at mechanical frequency $\omega_{m}$.}
	\label{fig:fig1}
\end{figure}

As depicted in Fig.~(\ref{fig:fig1}), we consider a hybrid system consisting of a cigar-shaped BEC of $N$ two-level atoms with mass $m_{a}$ and transition frequency $\omega_{a}$ trapped inside a membrane-in-the-middle optomechanical cavity with length $L$ where a mechanical oscillator (MO) which is free to oscillate at mechanical frequency $\omega_{m}$ has been placed on the right side of the BEC. The cavity is driven at rate $\eta=\sqrt{2\mathcal{P}\kappa/\hbar\omega_{p}}$ through the left mirror by a laser with frequency $\omega_{p}$, and wavenumber $k=\omega_{p}/c$ ($\mathcal{P}$ is the laser power and $\kappa$ is the cavity decay rate). Moreover,the cavity is injected by a broadband squeezed vacuum field with central frequency $ \omega_{sq} $ which is assumed to be at resonant with the cavity mode, i.e., $ \omega_{sq}=\omega_{0} $.

If the laser pump is far detuned from the atomic resonance, i.e., $\Delta_{a}=\omega_{p}-\omega_{a}\gg\gamma_{a}$ where $ \gamma_{a} $ is the atomic linewidth, then the excited electronic state of the atoms can be adiabatically eliminated and spontaneous emission can be neglected \cite{Masch Ritch 2004,Dom JB,Nagy Ritsch 2009}. In this way, the atomic wave function can be described by the scalar quantum field $ \Psi(x) $ and the Hamiltonian of the system in a frame rotating at the laser pump frequency can be written as
\begin{eqnarray}\label{H1}
H&=&\int_{-L_{0}/2}^{L_{0}/2} dx \Psi^{\dagger}(x)\Big[\frac{-\hbar^{2}}{2m_{a}}\frac{d^{2}}{dx^{2}}+\hbar U_{0} \cos^2(kx) a^{\dagger} a\nonumber\\
&&+\frac{1}{2} U_{s}\Psi^{\dagger}(x)\Psi(x)\Big] \Psi(x)-\hbar\Delta_{c} a^{\dagger} a + i\hbar\eta (a-a^{\dagger})\nonumber\\
&&+\frac{1}{2}\hbar\omega_{m}(p^{2}+q^{2})-\hbar\xi_{1} a^{\dagger}a q+\hbar\xi_{2} a^{\dagger}a q^2.
\end{eqnarray}

Here $L_{0}$ is the length of the BEC which has been extended symmetrically about the center of the cavity. It has also been assumed that the membrane reflectivity is so low that the cavity field can be considered as a single-mode field with annihilation operator\textbf{ $\hat a$} and central frequency $ \omega_{0}=n\pi c/L $ \cite{aliDCEJOSAB,bemani}. Besides, $ \Delta_{c}=\omega_{p}-\omega_{0} $ is the detuning between the cavity and the laser pump, $U_{0}=g_{0}^{2}/\Delta_{a}$ is the optical lattice barrier height per photon which represents the atomic backaction on the field, $g_{0}$ is the vacuum Rabi frequency, $U_{s}=4\pi\hbar^{2} a_{s}/m_{a}$ and $a_{s}$ is the two-body \textit{s}-wave scattering length \cite{Masch Ritch 2004,Dom JB}.

The first term in the last line of Eq.~(\ref{H1}) represents the free energy of the mechanical mode of the membrane, while the second and third terms describing the linear and quadratic  couplings of the mechanical mode of the membrane with the radiation pressure of the cavity with rates $\xi_{1} $ and  $\xi_{2} $, respectively. It should be noted that in a membrane-in-the-middle optomechanical system the coupling between the mechanical mode of the membrane and the optical field depends on the equilibrium position of the membrane relative to the nodes of the cavity mode \cite{Jayich}. This results in a cavity detuning which is a periodic function of the membrane equilibrium position whose Taylor expansion up to the second order leads to the linear ($\xi_{1}$) and the quadratic ($\xi_{2}$) couplings of the membrane to the radiation pressure of the cavity. Therefore, both the absolute value of the ratio $\xi_{2}/\xi_{1}$ and its sign can be controlled and manipulated experimentally through the equilibrium position of the membrane relative to the nodes of the cavity mode \cite{thompson}.

On the other hand, in the weakly interacting regime,where $U_{0}\langle a^{\dagger}a\rangle\leq 10\omega_{R}$ ($\omega_{R}=\hslash k^{2}/2m_{a}$ is the recoil frequency of the condensate atoms), and under the Bogoliubov approximation \cite{Nagy Ritsch 2009}, the atomic field operator can be expanded as the following single-mode quantum field
\begin{equation}\label{opaf}
\Psi(x)=\sqrt{\frac{N}{L_{0}}}+\sqrt{\frac{2}{L_{0}}}\cos(2kx) c,
\end{equation}
where the so-called Bogoliubov mode $ c $ corresponds to the quantum fluctuations of the atomic field about the classical condensate mode ($ \sqrt{\frac{N}{L_{0}}} $). As has been shown in Refs.~\cite{dalafi6, dalafi8} by substituting the atomic field operator of Eq.~(\ref{opaf}) into Eq.~(\ref{H1}), and defining the Bogoliubov mode quadratures as $Q=\chi (c+c^{\dagger})/\sqrt{2}$ and $P=(c-c^{\dagger})/\sqrt{2}i\chi$, where $ \chi=(\frac{4\omega_{R}+\frac{3}{2}\omega_{sw}}{4\omega_{R}+\frac{1}{2}\omega_{sw}})^{\frac{1}{4}} $ with $ \omega_{\textit{sw}}=8\pi \hbar a_{\textit{s}} N / (m_a L w^2)$ ($ w $ is the waist radius of the optical mode) being the \textit{s}-wave scattering frequency of the atomic collisions, the Hamiltonian of the system reduces to
\begin{eqnarray}\label{H}
H&=&\hbar\delta_{c}a^{\dagger}a+i\hbar\eta(a-a^{\dagger})+\frac{1}{2}\hbar\omega_{m}(p^{2}+q^{2})\nonumber\\
&&+\hbar a^{\dagger}a (\zeta Q+\xi_{2}q^2-\xi_{1}q)+\frac{1}{2}\hbar\omega_{c}(P^2+Q^2),
\end{eqnarray}
where $\delta_{c}=-\Delta_{c}+\frac{1}{2}N U_{0}$ is the effective Stark-shifted detuning. Based on Eq.~(\ref{H}), the Bogoliubov mode of the BEC behaves as a simple harmonic oscillator with frequency  $ \omega_{c}=\sqrt{(4\omega_{R}+\frac{1}{2}\omega_{sw})(4\omega_{R}+\frac{3}{2}\omega_{sw})} $  which couples to the radiation pressure of the optical field with the optomechanical strength $ \zeta=\frac{1}{2\chi}\sqrt{N}U_{0} $. Here, we have ignored the cross-Kerr nonlinear term of the BEC which is very weak compared to the radiation pressure interaction as has been shown in Ref.~\cite{dalafi7}.

\section{Dynamics of The System \label{Sec.III}}
The dynamics of the system is fully determined by the following set of nonlinear Heisenberg-Langevin equations:
\begin{subequations}\label{HaLaE}
\begin{eqnarray}
\dot{a}&=&-(i\delta_{c}+\kappa)a-\eta-i(\zeta Q-\xi_{1} q+\xi_{2} q^{2})a\nonumber\\
&&+\sqrt{2\kappa}\delta a_{in},\label{NHLa}\\
\dot{q}&=&\omega_{m} p,\label{NHLq}\\
\dot{p}&=&-\omega_{m} q+(\xi_{1}-2\xi_{2} q) a^{\dagger}a-\gamma_{m} p+\sqrt{2\gamma_{m}}\delta p_{in}\label{NHLp},\\
\dot{Q}&=&\omega_{c} P-\gamma_{c} Q+\sqrt{2\gamma_{c}} \delta Q_{in},\label{NHLc}\\
\dot{P}&=&-\omega_{c} Q-\gamma_{c} P-\zeta a^{\dagger}a+\sqrt{2\gamma_{c}} \delta P_{in},\label{NHLd}
\end{eqnarray}
\end{subequations}
where $ \gamma_{m} $ and $ \gamma_{c} $ are, respectively, the dissipation rates of the mechanical and the Bogoliubov modes. The system is affected by three uncorrelated quantum noise sources: the optical input vacuum noise, $ \delta a_{in} $, the Brownian noise $ \delta p_{in} $ acting on the moving membrane, and the quadrature noise operators $ \delta Q_{in} $ and $ \delta P_{in} $ acting on the Bogoliubove mode of the BEC arising from the harmonic trapping potential in which the BEC has been confined and also from the extra modes of the BEC which have been neglected in the single-mode approximation of Eq.~(\ref{opaf}) \cite{K Zhang, dalafi2, dalafi4}.

The Brownian noise acting on the membrane satisfies the Markovian correlation function $ \langle\delta p_{in}(t)\delta p_{in}(t^{\prime})\rangle=(n_{m}+1/2)\delta(t-t^{\prime}) $ with $ n_{m}=[\exp(\hbar\omega_{m}/k_{B}T)-1]^{-1} $ as the mean number of thermal phonons of the mechanical mode \cite{giovan}. Moreover, the input noises of the BEC satisfy similar correlation functions $ \langle\delta Q_{in}(t)\delta Q_{in}(t^{\prime})\rangle=\langle\delta P_{in}(t)\delta P_{in}(t^{\prime})\rangle=(n_{c}+1/2)\delta(t-t^{\prime}) $, $ \langle\delta Q_{in}(t)\delta P_{in}(t^{\prime})\rangle=-\langle\delta P_{in}(t)\delta Q_{in}(t^{\prime})\rangle=i\delta(t-t^{\prime})/2 $ with $ n_{c}=[\exp(\hbar\omega_{c}/k_{B}T)-1]^{-1} $ as the mean number of thermal excitations of the Bogoliubov mode of the BEC \cite{K Zhang}.

The quantum vacuum fluctuations of the optical field in the absence of the squeezed vacuum injection satisfy the Markovian correlation functions $ \langle\delta a_{in}(t)\delta a_{in}^{\dagger}(t^{\prime}\rangle)=\delta(t-t^{\prime}) $ and $ \langle\delta a_{in}^{\dagger}(t)\delta a_{in}(t^{\prime}\rangle)=0 $. However, if the cavity is driven by a broadband squeezed vacuum field with a spectrum centered at the cavity resonance frequency $ \omega_{sq}=\omega_{0} $ then the optical noise correlation functions satisfy the Markovian relations $ \langle\delta a_{in}(t)\delta a_{in}(t^{\prime}\rangle)=M_{s}\delta(t-t^{\prime}) $ and $ \langle\delta a_{in}^{\dagger}(t)\delta a_{in}(t^{\prime}\rangle)=N_{s}\delta(t-t^{\prime}) $ where in the case of pure squeezing $ M_{s}=(1/2)\sinh(2r)\exp(i\phi) $ and $ N_{s}=\sinh^2(r) $ with $ r $ and $ \phi $ being, respectively, the strength and the phase of squeezing, so that $ |M_{s}|^2=N_{s}(N_{s}+1) $ \cite{sq1, sq2, motazedi}.

Now, in order to find the solutions to the set of nonlinear equations (\ref{NHLa}-\ref{NHLd}) we decompose each operator as the sum of its mean-field value and a small fluctuation around it. By substituting $ a=\alpha+\delta a $ for the optical field, and $ o=\bar{o}+\delta o $ with $ o=q,p,Q,P $ for the quadratures of the mechanical and atomic fields into Eqs.~(\ref{NHLa}-\ref{NHLd}) one can obtain the following set of nonlinear first-order ordinary differential equations for the classical mean-fields
\begin{subequations}\label{mv}
\begin{eqnarray}
\dot{\alpha}&=&-(i\Delta+\kappa)\alpha-\eta,\\
\dot{\bar{q}}&=&-\omega_{m}\bar{p},\\
\dot{ \bar{p}}&=&-(\omega_{m}+2\xi_{2}|\alpha|^2)\bar{q}+\xi_{1}|\alpha|^2-\gamma_{m}\bar{p},\\
\dot{\bar{Q}}&=&\omega_{c}\bar{P}-\gamma_{c}\bar{Q},\\
\dot{\bar{P}}&=&-\omega_{c}\bar{Q}-\zeta |\alpha|^2-\gamma_{c}\bar{P},
\end{eqnarray}
\end{subequations}
where $ \Delta=\delta_{c}+\zeta\bar{Q}-\xi_{1}\bar{q}+\xi_{2}\bar{q}^2 $ is the effective detuning of the cavity. By defining the optical quadrature fluctuations as $ \delta X=\frac{1}{\sqrt{2}}(\delta a+\delta a^{\dagger}) $ and $ \delta Y=\frac{1}{\sqrt{2}i}(\delta a-\delta a^{\dagger}) $ the linearized QLEs are obtained in the following compact matrix form:
\begin{equation}\label{nA}
\delta\dot{u}(t)=A\delta u(t)+\delta n(t),
\end{equation}
where $ \delta u=[\delta X, \delta Y,\delta q, \delta p, \delta Q, \delta P ]^{T}  $ is the vector of continuous variable fluctuation operators and $ \delta n=[\sqrt{2\kappa}\delta X_{in}, \sqrt{2\kappa}\delta Y_{in}, 0, \sqrt{2\gamma_{m}}\delta p_{in},\sqrt{2\gamma_{c}}\delta Q_{in}, \sqrt{2\gamma_{c}}\delta P_{in}]^{T} $ is the corresponding vector of noises in which $ \delta X_{in}=\frac{1}{\sqrt{2}}(\delta a_{in}+\delta a_{in}^{\dagger}) $ and $ \delta Y_{in}=\frac{1}{\sqrt{2}i}(\delta a_{in}-\delta a_{in}^{\dagger}) $ are the input noise quadratures of the optical field . The $ 6\times 6 $ drift matrix $ A $ is given by
\begin{equation}
A=\left(\begin{array}{cccccc}
-\kappa & \Delta & -\sqrt{2}\alpha_{I}\beta & 0 & \sqrt{2}\alpha_{I}\zeta & 0 \\
-\Delta & -\kappa & \sqrt{2}\alpha_{R}\beta & 0 & -\sqrt{2}\alpha_{R}\zeta & 0 \\
0 & 0 & 0 & \omega_{m} & 0 & 0 \\
\sqrt{2}\alpha_{R}\beta & \sqrt{2}\alpha_{I}\beta & -\omega_{b} & -\gamma_{m} & 0 & 0 \\
0 & 0 & 0 & 0 & -\gamma_{c} & \omega_{c} \\
-\sqrt{2}\alpha_{R}\zeta & -\sqrt{2}\alpha_{I}\zeta & 0 & 0 & -\omega_{c} & -\gamma_{c} \\
  \end{array}\right),
\label{A}
\end{equation}
where $ \beta=\xi_{1}-2\xi_{2}\bar{q} $, $ \omega_{b}=\omega_{m}+2\xi_{2}|\alpha|^2 $, and $ \alpha_{R} $ and $ \alpha_{I} $ are, respectively, the real and imaginary parts of the optical mean field. In the following two sections, we will investigate how the QOC affects the dynamical behavior of the mean-fields and also the multistability of the system in the stationary state.

\section{mean-fields dynamics \label{Sec.IV}}
If the damping rate of the cavity is faster than those of the mechanical and Bogoliubov modes, i.e. $\kappa\gg\gamma_{m}, \gamma_{c}$, the optical field follows the dynamics of the mechanical and atomic oscillators adiabatically. Therefore, in the so-called adiabatic approximation Eq.~(\ref{mv}a) leads to the following equation
\begin{equation}\label{malpha}
\alpha(t)\approx\frac{-\eta}{i\Big(\delta_{c}+\zeta\bar{Q}(t)-\xi_{1}\bar{q}(t)+\xi_{2}\bar{q}^2(t)\Big)+\kappa}.
\end{equation}

Therefore, Eqs.~(\ref{mv}b-\ref{mv}e) together with Eq.~(\ref{malpha}) lead to the following set of second-order nonlinear ordinary differential equations for the mean-fields of the quadratures $q$ and $Q$ of the mechanical and the Bogoliubov modes, respectively, which are coupled to each other through the mean-value of the optical field
\begin{subequations}\label{qQ}
\begin{eqnarray}
\ddot{\bar{q}}(t)+\gamma_{m}\dot{\bar{q}}(t)+\tilde{\omega}_{m}^{2} \bar{q}(t)&=&\xi_{1}\omega_{m}|\alpha(t)|^2,\\
\ddot{\bar{Q}}(t)+2\gamma_{c}\dot{\bar{Q}}(t)+\tilde{\omega}_{c}^{2}\bar{Q}(t)&=&-\zeta\omega_{c}|\alpha(t)|^2,
\end{eqnarray}
\end{subequations}
where $\tilde{\omega}_{m}=\sqrt{\omega_{m}(\omega_{m}+2\xi_{2}|\alpha(t)|^2)}$ and $\tilde{\omega}_{c}=\sqrt{\omega_{c}^{2}+\gamma_{c}^{2}}$ are, respectively, the effective frequencies of the mechanical and the Bogoliubov modes. As is seen, the effective frequency of the mechanical mode has an explicit dependence on the mean value of the optical field through the parameter $\xi_{2}$ of the QOC. On the other hand, since the Bogoliubov mode is coupled indirectly to the mechanical mode through the mediation of the optical field, its effective frequency is also modified by the QOC of the mechanical mode, as will be shown in the following.

The set of Eqs.~(\ref{qQ}a), (\ref{qQ}b) and (\ref{malpha}) can be solved numerically. Here, we analyze our results based on the experimentally feasible parameters given in Refs.~\cite{Brenn Science, Ritter Appl. Phys. B}. We consider an optical cavity of length $ L=178 \mu$m whose bare frequency is $ \omega_{0}=2.41494\times 10^{15} $Hz corresponding to a wavelength of $ \lambda=780 $nm. The membrane oscillates with frequency $ \omega_{m}=10^5 $Hz and has a damping rate $ \gamma_{m}=2\pi\times 100 $Hz. The cavity contains  $ N=10^5 $ Rb atoms and is coherently driven at amplitude $ \eta $ by a pump laser with frequency $ \omega_{p} $ through the left end mirror and has a damping rate of $ \kappa=2\pi\times 1.3 $MHz. The atomic $ D_{2} $ transition corresponding to the atomic transition frequency $ \omega_{a}=2.41419\times 10^{15} $Hz couples to the mentioned mode of the cavity. The atom-field coupling strength is $ g_{0}=2\pi\times 14.1 $MHz and the recoil frequency of the atoms is $ \omega_{R}=23.7 $KHz.

\begin{figure}
\centering
\includegraphics[width=4.27cm]{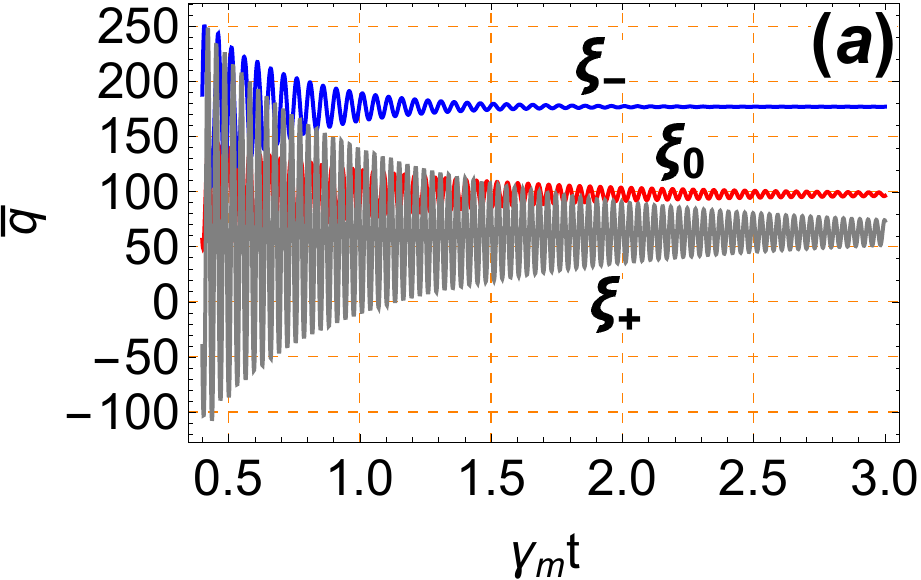}
\includegraphics[width=4.29cm]{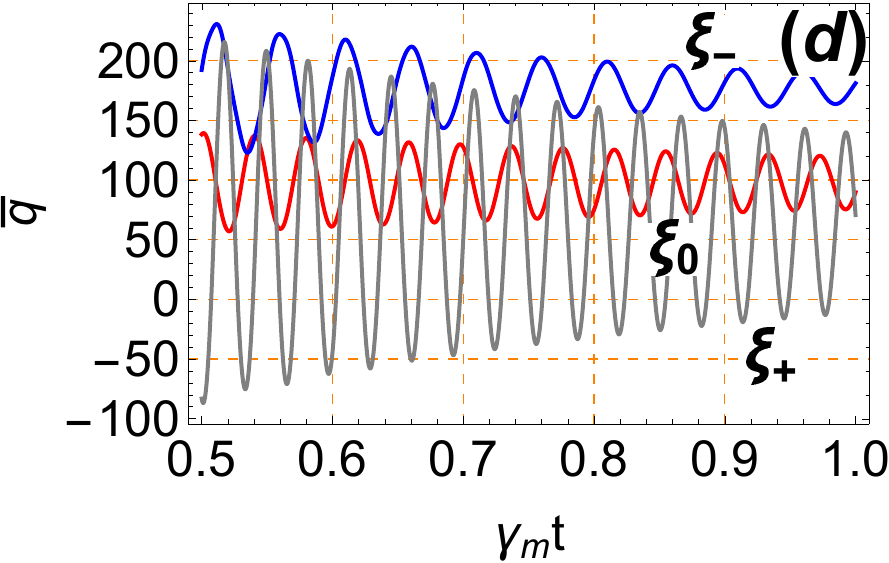}
\includegraphics[width=4.27cm]{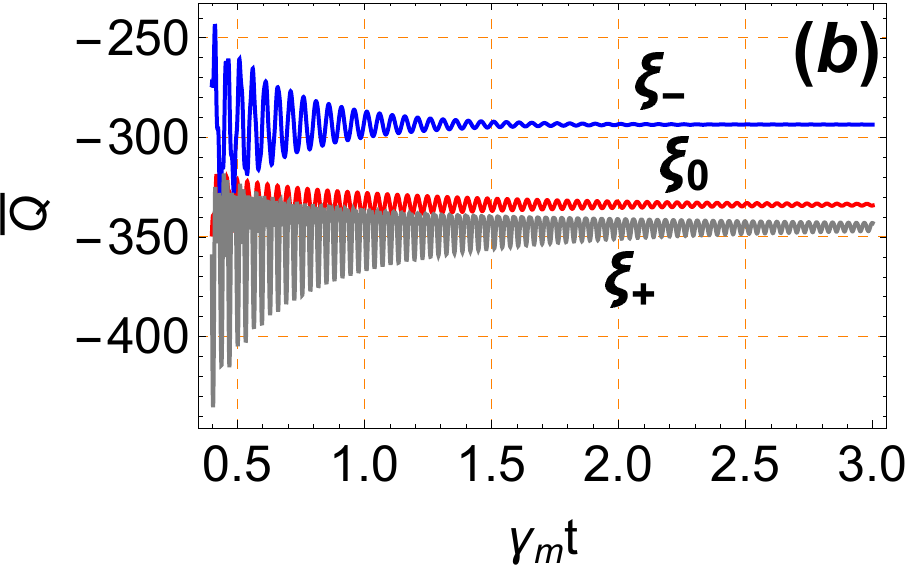}
\includegraphics[width=4.3cm]{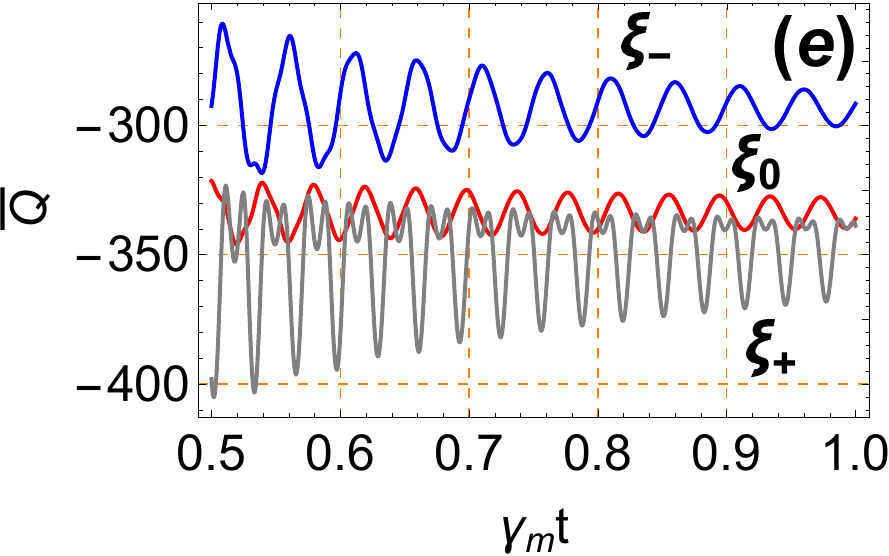}
\includegraphics[width=4.15cm]{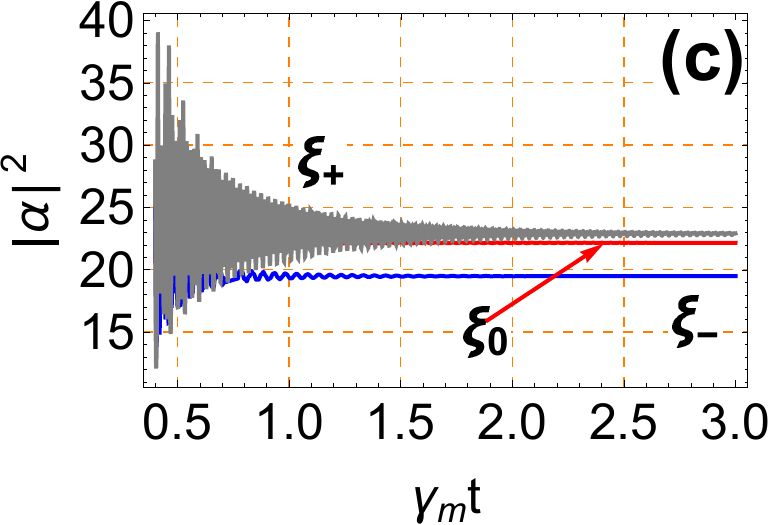}
\includegraphics[width=4.15cm]{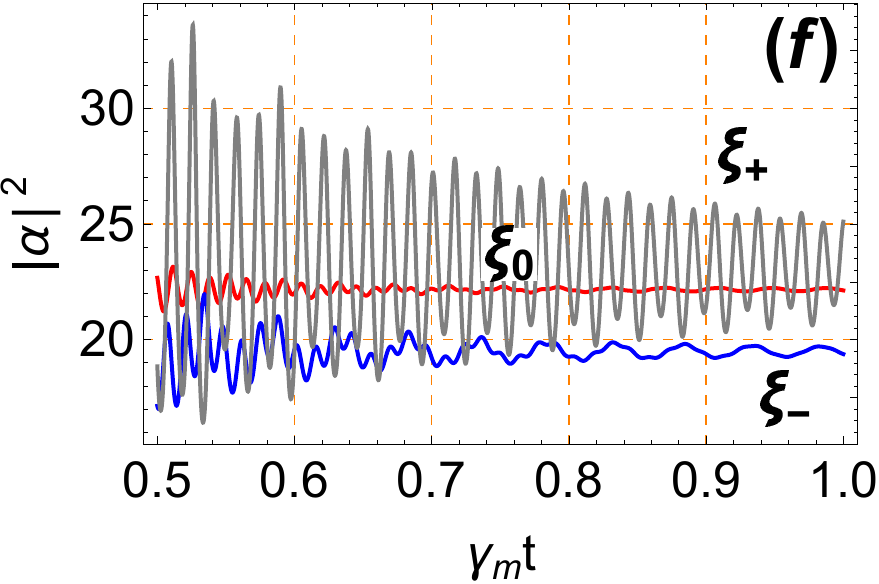}
\caption{
(Color online) The mean values of (a) the \textit{q}-quadrature of the mechanical mode, (b) the \textit{Q}-quadrature of the Bogoliubov mode of the BEC, and (c) the mean number of cavity photons versus the normalized time $\gamma_{m}t$ when the cavity is pumped at rate $\eta=100\kappa$ and the effective detuning has been fixed at $\delta_{c}=50\kappa$ for $\xi_{2}=0$ (the line indicated by $\xi_{0}$), $\xi_{2}=-0.003\xi_{1}$ (the line indicated by $\xi_{-}$) and $\xi_{2}=+0.003\xi_{1}$ (the line indicated by $\xi_{+}$). The cavity contains $ N=10^5 $ Rb atoms and has a length of $ L=178 \mu$m. The other parameters are $\kappa=2\pi\times1.3$MHz, $\lambda=780$nm, $\xi_{1}=0.05\kappa$, $\gamma_{m}=2\pi\times 100$Hz, $\gamma_{c}=0.001\kappa$ and $\omega_{sw}=0.5\omega_{R}$. Panels (d)-(f) show the temporal behaviors of the same mean fields in a shorter time interval.}
\label{fig:fig2}
\end{figure}

In Fig.~(\ref{fig:fig2}) we have shown the temporal behavior of the mean values of the \textit{q}-quadrature of the mechanical mode [Fig.~\ref{fig:fig2}(a)] and the \textit{Q}-quadrature of the Bogoliubov mode [Fig.~\ref{fig:fig2}(b)] as well as the optical mean-field [Fig.~\ref{fig:fig2}(c)] when the cavity is pumped at rate $\eta=100\kappa$ and the effective detuning has been fixed at $\delta_{c}=50\kappa$ for three different values of QOC parameter $\xi_{2}=0$ (the line indicated by $\xi_{0}$), $\xi_{2}=-0.003\xi_{1}$ (the line indicated by $\xi_{-}$), and $\xi_{2}=+0.003\xi_{1}$ (the line indicated by $\xi_{+}$) with $ \xi_{1}=0.05\kappa $. In Figs.~\ref{fig:fig2}(d)-\ref{fig:fig2}(f), the mean-fields have been plotted in a shorter time interval for the convenience of comparing their oscillatory behaviors.

As is seen from Fig.~(\ref{fig:fig2}), the overall effect of the QOC manifests as the increase in the amplitude of oscillations. Besides,  the presence of the QOC with a positive sign increases while with a negative sign decreases the oscillation frequencies of both the mechanical and the Bogoliubov modes compared to those in its absence. It is due to the dependence of the effective frequency of the mechanical mode to the optical mean-field through $\xi_{2}$, i.e., $\tilde{\omega}_{m}=\sqrt{\omega_{m}(\omega_{m}+2\xi_{2}|\alpha|^2)}$ and the coupling of the Bogoliubov mode to the mechanical mode through the mediation of the optical mode. Moreover, the presence of QOC with a positive sign increases while with a negative sign decreases the relaxation times of oscillations of both the mechanical and the Bogoliubov modes compared to those in its absence.

The results obtained here for a hybrid optomechanical system are comparable to those of Ref.~\cite{wang} where a similar behavior has been observed for the mechanical frequency of a bare optomechanical cavity. The  interesting point is that in the hybrid system the Bogoliubov mode of the BEC behaves as a secondary mechanical mode which is coupled indirectly to the mechanical mode of the membrane with the mediation of the optical field. Therefore, the variation of the mechanical frequency due to the QOC to the optical field is also transferred indirectly to the Bogoliubov mode which leads to the modification of its frequency.

\section{optical multistability \label{Sec.V}}
In this section we investigate the effects of the QOC on the multistability behavior of the system. For this purpose, we should find the stationary-state solutions of the mean field equations [Eqs.~(\ref{mv}a)-(\ref{mv}e)]. In the steady-state, the mean-field of the optical mode is given by $ |\alpha|^2=\eta^2/(\Delta^2+\kappa^2) $ while those of the mechanical and Bogoliubov modes are, respectively, $\bar{q}=\xi_{1}|\alpha|^2/(\omega_{m}+2\xi_{2}|\alpha|^2)$, $\bar{p}=0$, $\bar{Q}=-(\gamma_{c}/\omega_{c})\bar{P}-(\zeta/\omega_{c})|\alpha|^{2}$, and $\bar{P}=(\gamma_{c}/\omega_{c})\bar{Q}$. By solving this set of nonlinear algebraic equations, we obtain the steady-state value of the optical mode in terms of the effective detuning $ \delta_{c} $ for different values of the QOC parameter $ \xi_{2} $.

\begin{figure}
	\centering
	\includegraphics[width=1.68in]{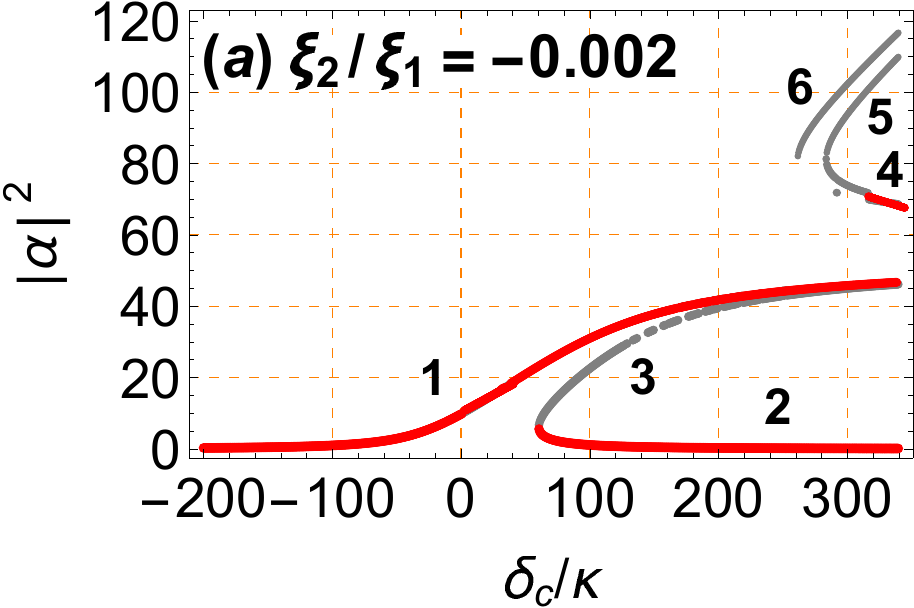}
	\includegraphics[width=1.65in]{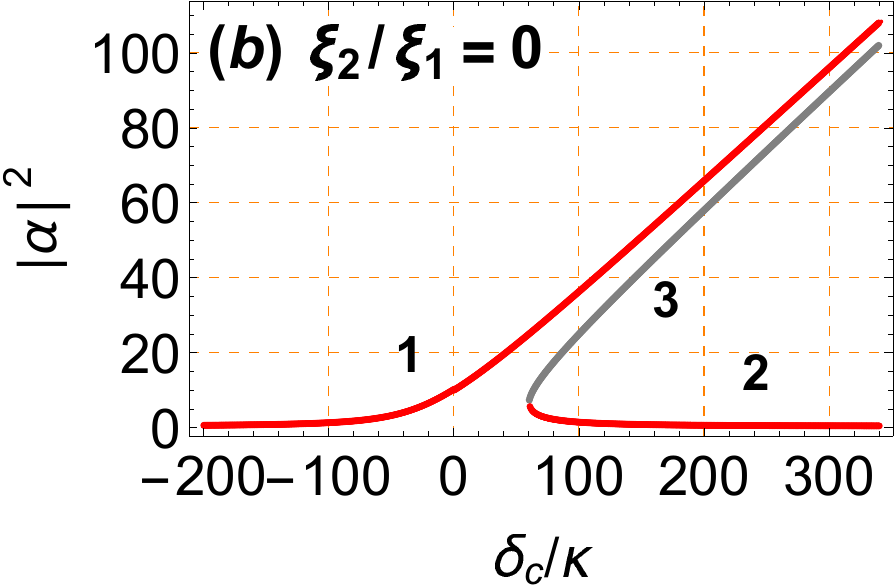}
	\includegraphics[width=1.68in]{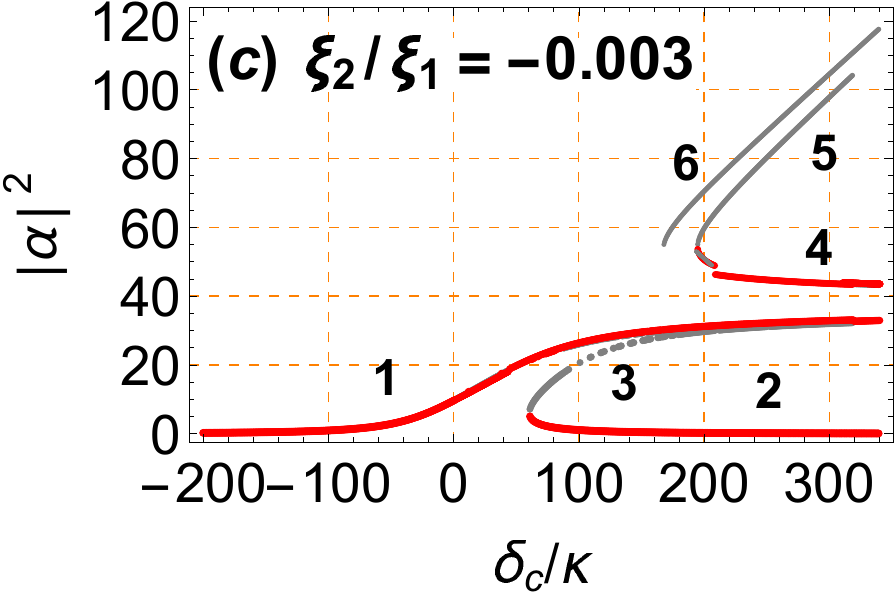}
	\includegraphics[width=1.68in]{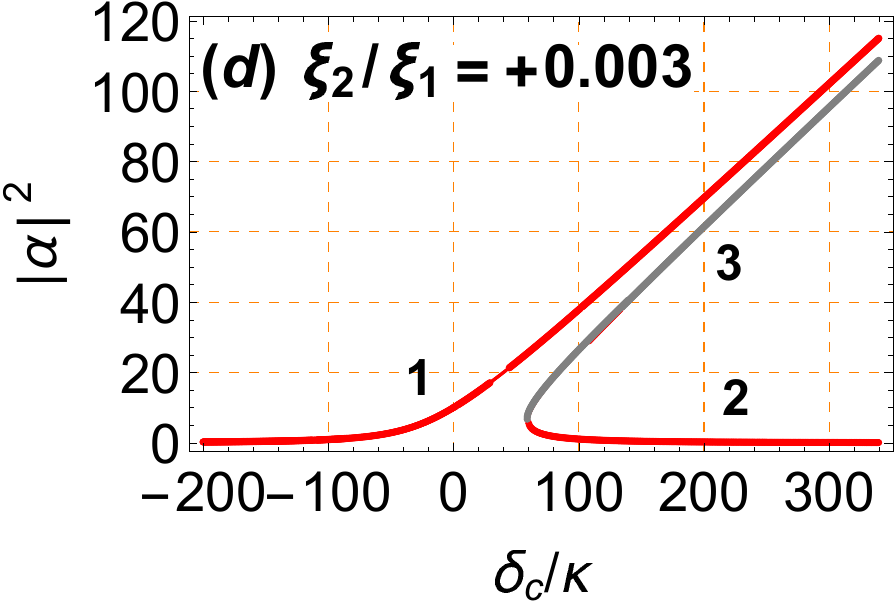}
	\includegraphics[width=1.68in]{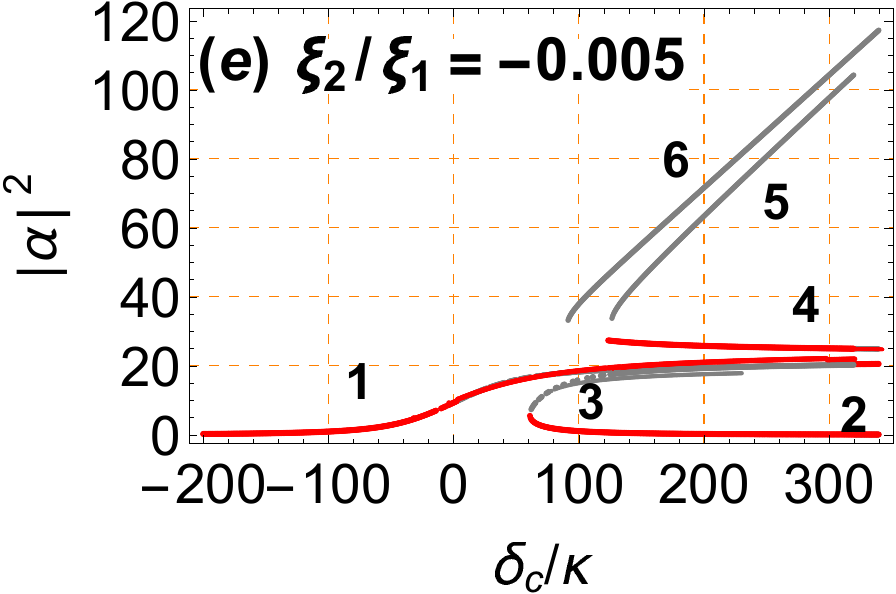}
	\includegraphics[width=1.68in]{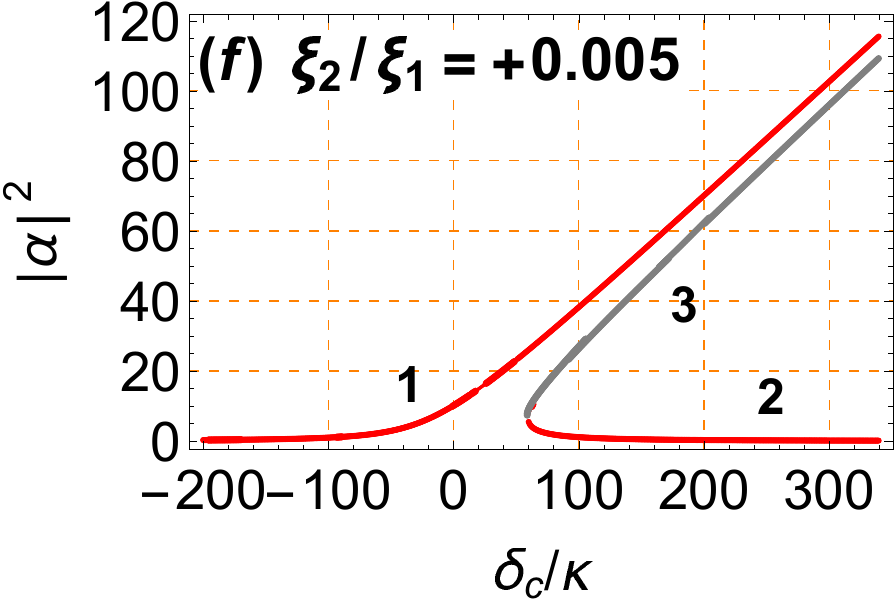}
	\caption{
		(Color online) The mean number of cavity photons versus the normalized effective detuning $ \delta_{c}/\kappa $ for $ \xi_{2}=-0.002\xi_{1}, 0 $ (panels a and b), $ \xi_{2}=\pm0.003\xi_{1} $ (panels c and d), and $\xi_{2}=\pm0.005\xi_{1} $ (panels e and f)  when the cavity is driven at rate $\eta=100\kappa$. Each numbered curve corresponds to the stable or unstable solution for the steady-state mean value of the optical field (see the text for details). The other parameters are the same as those of Fig.~\ref{fig:fig2}.
	}
	\label{fig:fig3}
\end{figure}

In Fig.~(\ref{fig:fig3}) we have demonstrated the steady-state mean value of the optical field (the mean number of cavity photons) versus the normalized effective detuning $ \delta_{c}/\kappa $ in the absence of the QOC [Fig.~\ref{fig:fig3}(b)] and in its presence with $\xi_{2}/\xi_{1}=-0.002$ [Fig.~\ref{fig:fig3}(a)], $\xi_{2}/\xi_{1}=\pm 0.003$ [Figs.~\ref{fig:fig3}(c),\ref{fig:fig3}(d)], and  $\xi_{2}/\xi_{1}=\pm 0.005$ [Figs.~\ref{fig:fig3}(e),\ref{fig:fig3}(f)]. As is seen from Fig.~\ref{fig:fig3}(b) in the absence of the QOC, i.e., when $ \xi_{2}=0 $ the system shows an ordinary bistability behavior for $ \delta_{c}>60\kappa $. Here, the system is stable along the branches 1 and 2 while it is unstable along the branch 3. The stability conditions can be obtained, for example, by using the Routh-Hurwitz criteria \cite{RH}. In fact, the system is stable only if all the eigenvalues of the drift matrix $ A $ [Eq.~(\ref{A})] have negative real parts.

In order to see how the QOC with a negative sign affects the multistability behavior of the system, we have shown in Figs.~\ref{fig:fig3}(a),\ref{fig:fig3}(c), and \ref{fig:fig3}(e) the optical mean-field value versus the normalized effective detuning $ \delta_{c}/\kappa $ for three negative values of  $ \xi_{2}/\xi_{1} $. Based on our numerical calculations the system is stable along the branches 1, 2 and 4 of Figs.~\ref{fig:fig3}(a),\ref{fig:fig3}(c), and \ref{fig:fig3}(e) while it is unstable along the branches 3, 5 and 6. Here, the system shows a single stability behavior for $\delta_{c}<60\kappa$. However, for $\delta_{c}>60\kappa$ it shows a bistability behavior for a limited interval and afterwards it switches to a tristability behavior. Interestingly, the transition threshold from bistability to tristability is shifted to the lower values of $\delta_{c}$ for larger (negative) values of $|\xi_{2}|$. Therefore, the mentioned threshold can be controlled by the absolute value of the QOC. Such a bistability-tristability transition can be used for all-optical switching purposes functioning as memory devices for optical computing and quantum information processing \cite{shahidani, sheng}.

On the other hand, for positive values of the QOC, as shown in Fig.~\ref{fig:fig3}(d) with $ \xi_{2}/\xi_{1}=+0.003 $ and Fig.~\ref{fig:fig3}(f) with $ \xi_{2}/\xi_{1}=+0.005 $,  the system shows a bistability behavior very similar to that in the absence of the QOC [Fig.~\ref{fig:fig3}(b)] with the difference that here the mean number of photons is increased a little bit for each value of the effective detuning for larger values of QOC. In Figs.~\ref{fig:fig3}(d) and \ref{fig:fig3}(f) the system is stable along the branches 1 and 2 while it is unstable along the branch 3. Therefore, one can control the optical multistability behavior of such a hybrid system through the sing of the QOC between the mechanical and the optical modes. Moreover, for negative values of QOC one can shift the  threshold of the optical bistability-tristability transition through the absolute value of the QOC.

\section{steady-state values of quantum fluctuations \label{Sec.VI}}
In order to examine the effect of the squeezed vacuum injection on the optomechanical properties of the system, we should study the quantum fluctuations of the system. Here, we exploit the injection of a broadband squeezed vacuum light in order to control the squeezing degrees of the atomic and mechanical modes and their bipartite entanglement when the system reaches the stationary state. In addition, we investigate the impact of the QOC on the quantum fluctuations of the system. 

Due to the linearized dynamics of the fluctuations and since all noises are Gaussian the steady state is a zero-mean Gaussian state which is fully characterized by the $6\times6$ stationary correlation matrix (CM) V , with components $V_{ij}=[\langle \delta u_{i}(\infty)\delta u_{j}(\infty)+\delta u_{j}(\infty)\delta u_{i}(\infty)\rangle]/2$. In order to obtain $\delta u_{i}(\infty)$ one needs to solve Eq.~(\ref{nA}) whose solution is given by
\begin{equation}\label{ut}
\delta u(t)=\mathcal{M}(t)\delta u(0)+\int_{0}^{t}dt'\mathcal{M}(t')\delta n(t-t'),
\end{equation}
where  $\mathcal{M}(t)=\exp(A t)$. If the stability conditions are satisfied then $\mathcal{M}(\infty)=0$ and therefore the steady-state solution is obtained as
\begin{equation}\label{uinf}
\delta u_{i}(\infty)=\int_{0}^{\infty}dt' \sum_{k} \mathcal{M}_{ik}(t')\delta n_{k}(t-t').
\end{equation}
Using Eq.(\ref{uinf}) the steady-state values of the CM elements are obtained as
\begin{equation}\label{Vij}
V_{ij}=\sum_{k,l}\int_{0}^{\infty}dt\int_{0}^{\infty}dt' \mathcal{M}_{ik}(t)\mathcal{M}_{jl}(t') D_{kl}(t-t'),
\end{equation}
where $D(t-t^{\prime})$ is the diffusion matrix whose elements are obtained by the relation 
\begin{eqnarray}\label{Dt}
D_{kl}(t-t^{\prime})&=&\frac{1}{2}\langle\delta n_{k}(t)\delta n_{l}(t^{\prime})+\delta n_{l}(t^{\prime})\delta n_{k}(t)\rangle,\nonumber\\
&&=D_{kl}\delta(t-t^{\prime}).
\end{eqnarray}
By substituting Eq.~(\ref{Dt}) into Eq.~(\ref{Vij}) the CM in the stationary state is obtained as
\begin{equation}\label{Vst}
V=\int_{0}^{\infty}dt^{\prime} \mathcal{M}(t^{\prime}) D \mathcal{M}^{T}(t').
\end{equation}
If the stability conditions are satisfied so that $\mathcal{M}(\infty)=0$ then Eq.~(\ref{Vst}) leads to the Lyapunov equation
\begin{equation}\label{Lyap}
A V+ V A^{T}=-D.
\end{equation}

Based on Eq.~(\ref{Dt}) and using the quantum noise correlation functions of the system explained in Sec.~\ref{Sec.III}, we can determine the diffusion matrix. In the absence of the squeezed vacuum light injection the diffusion matrix $D$ is a diagonal matrix given by $D=diag[\kappa,\kappa,0,\gamma_{m}^{\prime},\gamma_{c}^{\prime},\gamma_{c}^{\prime}$], while in the presence of injected squeezing it has the following form
\begin{equation}D=\left(\begin{array}{cccccc}
2\kappa(M_{s}^{(R)}+N_{s}+\frac{1}{2}) & 2\kappa M_{s}^{(I)} & 0 & 0 & 0 & 0 \\
2\kappa M_{s}^{(I)} & 2\kappa(M_{S}^{(R)}+N_{s}+\frac{1}{2}) & 0 & 0 & 0 & 0 \\
0 & 0 & 0 & 0 & 0 & 0 \\
0 & 0 & 0 & \gamma_{m}^{\prime} & 0 & 0 \\
0 & 0 & 0 & 0 & \gamma_{c}^{\prime} & 0 \\
0 & 0 & 0 & 0 & 0 & \gamma_{c}^{\prime} \\
  \end{array}\right),
\label{D}
\end{equation}
where $M_{s}^{(R)}$ and $M_{s}^{(I)}$ are, respectively, the real and imaginary parts of the squeezed vacuum parameter $ M_{s} $ while $\gamma_{m}^{\prime}=\gamma_{m}(2n_{m}+1)$ and $\gamma_{c}^{\prime}=\gamma_{c}(2n_{c}+1)$.

In order to see when a quantum oscillator exhibits quadrature squeezing we should consider a measure for the degree of squeezing. If $\delta x$ and $\delta y$ are the quadratures of a quantum oscillator (like those of the mechanical mode of the membrane or the Bogoliubov mode of the BEC) which satisfy the commutation relation $[\delta x, \delta y]=i$, the Heisenberg uncertainty relation is given by $\Delta x \Delta y\geq\sigma_{ZPF}$ where $\sigma_{ZPF}=|\langle[\delta x,\delta y]\rangle|/2$ is the zero-point fluctuation and $\Delta x=\sqrt{\langle\delta x^2\rangle-\langle\delta x\rangle^{2} }$ and $\Delta y=\sqrt{\langle\delta y^2\rangle-\langle\delta y\rangle^{2} }$ are quadrature uncertainties. If the quantum oscillator is in a state in which one of the variances $\sigma_{x}=(\Delta x)^2$ or $\sigma_{y}=(\Delta y)^2$ is less than $\sigma_{ZPF}=1/2$ then the corresponding state is a squeezed one. One of the most convenient measures for the degree of squeezing is the definition in the dB unit which is given by $-10\log_{10}(\sigma_{j}/\sigma_{ZPF})$ for $j=x$ or $y$.
Based on this definition, if the oscillator is in a state in which one of the variances is $\sigma_{j}=\sigma_{ZPF}/2$ which corresponds to $50\%$ noise reduction  below the zero-point level, the degree of squeezing is approximately 3 dB for the corresponding quadrature.

On the other hand, the bipartite entanglement can be calculated by using the logarithmic negativity \cite{eis}:
\begin{equation}\label{en}
E_N=\mathrm{max}[0,-\mathrm{ln} 2 \eta^-],
\end{equation}
where  $\eta^{-}\equiv2^{-1/2}\left[\Sigma(\mathcal{V}_{bp})-\sqrt{\Sigma(\mathcal{V}_{bp)}^2-4 \mathrm{det} \mathcal{V}_{bp}}\right]^{1/2}$  is the lowest symplectic eigenvalue of the partial transpose of the $4\times4$ CM, $\mathcal{V}_{bp}$, associated with the selected bipartition, obtained by neglecting the rows and columns of the uninteresting modes
\begin{equation}\label{bp}
\mathcal{V}_{bp}=\left(
     \begin{array}{cc}
     \mathcal{B}&\mathcal{C}\\
      \mathcal{C}^{T}&\mathcal{B}^{\prime}\\
       \end{array}
   \right),
\end{equation}
and $\Sigma(\mathcal{V}_{bp})=\mathrm{det} \mathcal{B}+\mathrm{det} \mathcal{B}^{\prime}-2\mathrm{det} \mathcal{C}$.

Based on our numerical calculations, the squeezing does not occur in the \textit{p}-(\textit{P}-)quadrature of the mechanical (atomic) mode while the conjugate quadrature \textit{q}(\textit{Q}) shows squeezing behavior. Here, the variances $\sigma_{q}=\langle\delta q^2\rangle$ and $\sigma_{Q}=\langle\delta Q^2\rangle$  can be determined in terms of the CM elements so that for the mechanical and the Bogoliubov modes we have, respectively, $\sigma_{q}=V_{33}$ and $\sigma_{Q}=V_{55}$.

\begin{figure}
 	\centering
 	\includegraphics[width=1.65in]{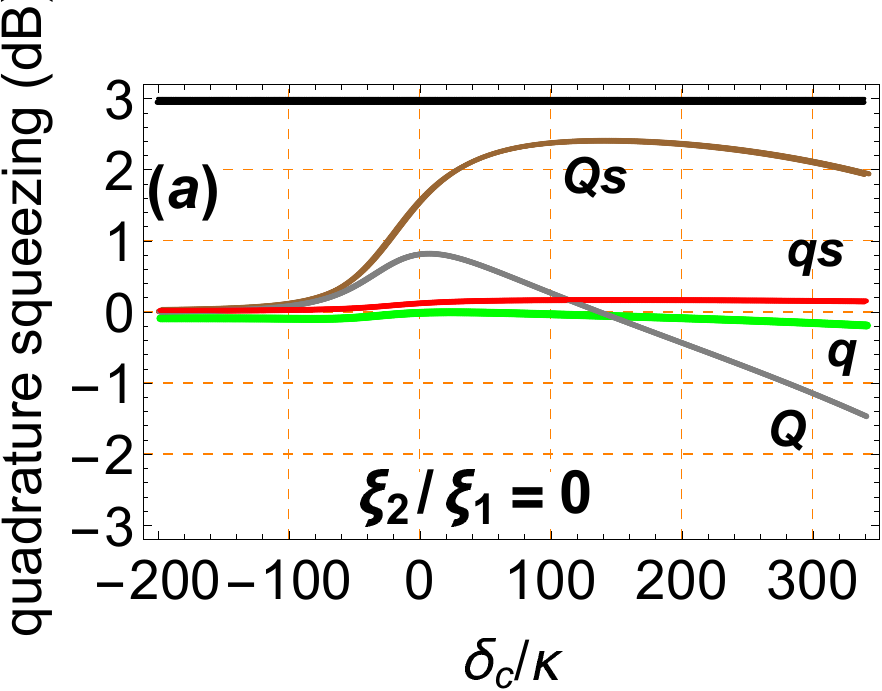}
 	\includegraphics[width=1.69in]{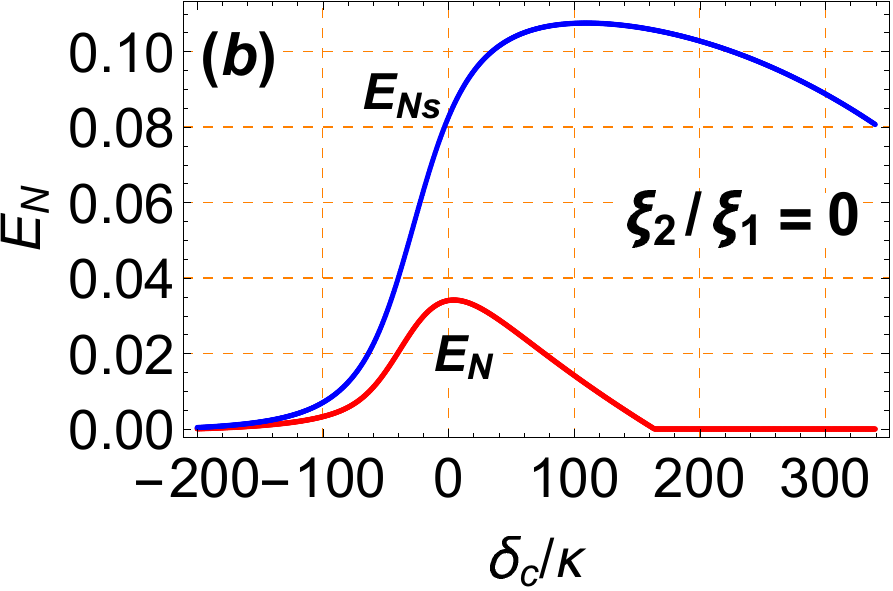}
 	\includegraphics[width=1.67in]{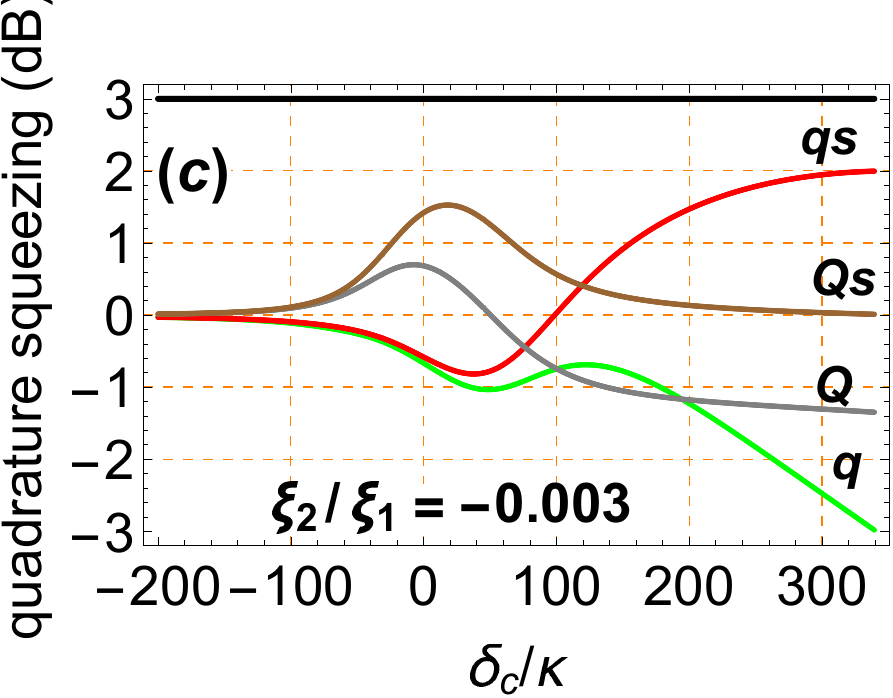}
 	\includegraphics[width=1.67in]{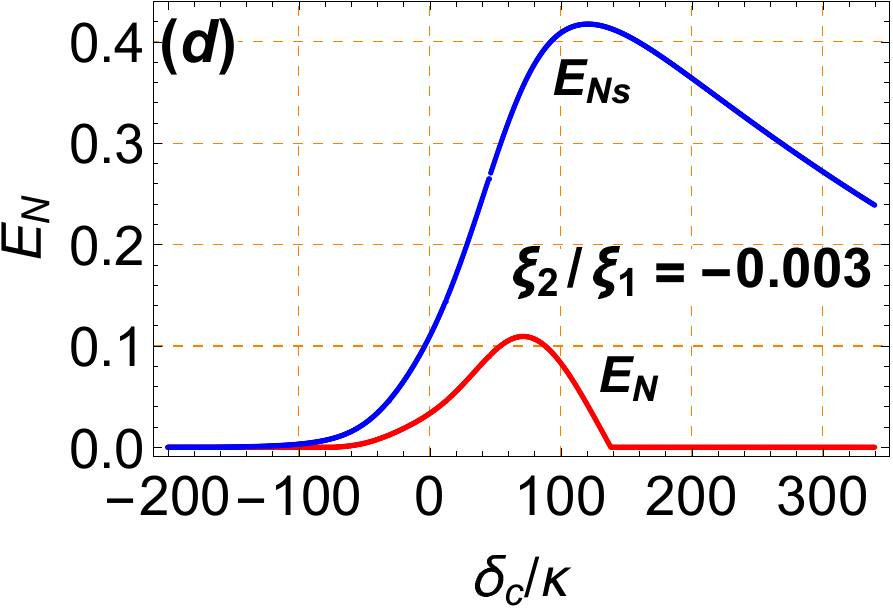}
 	\includegraphics[width=1.67in]{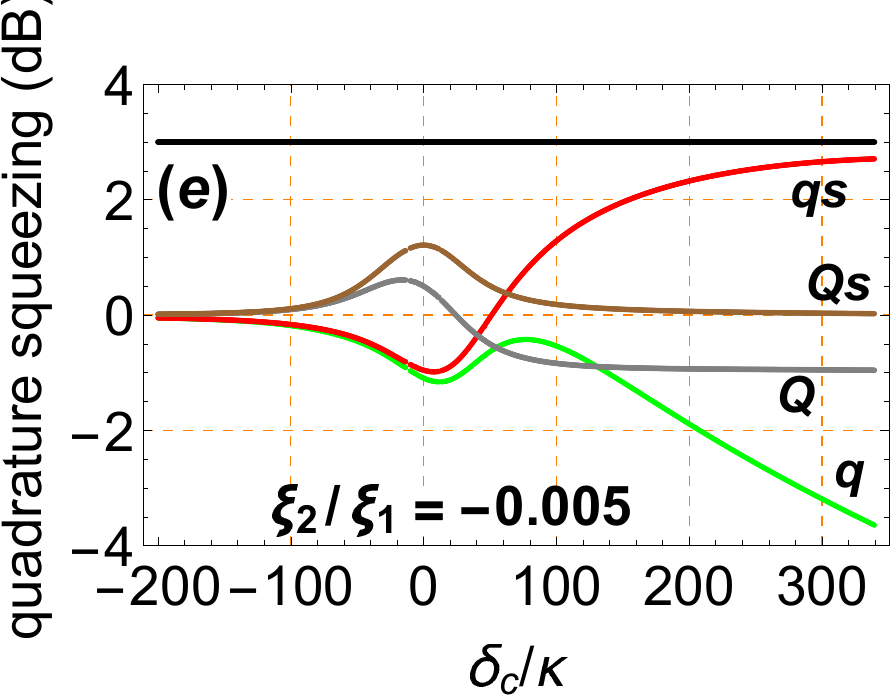}
 	\includegraphics[width=1.67in]{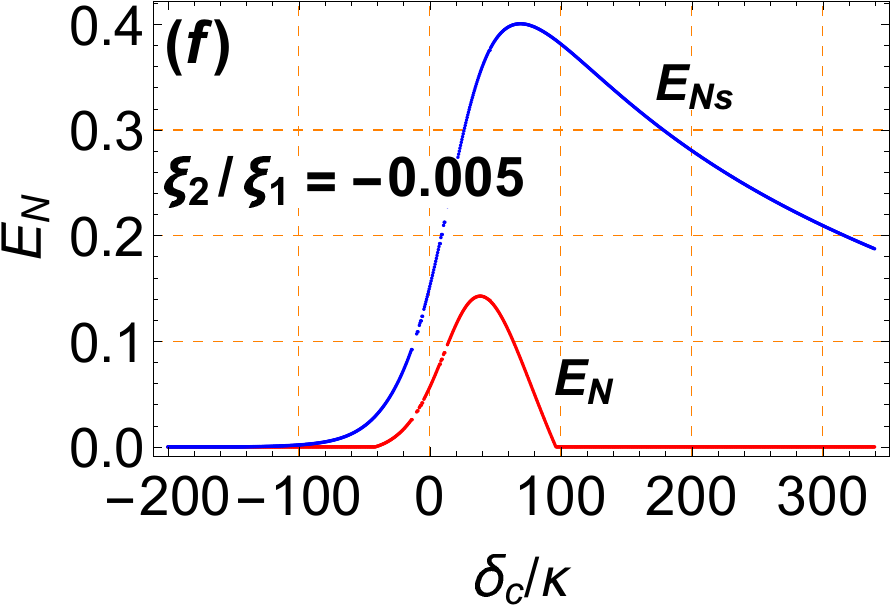}
 	\caption{
 		(Color online) The quadrature squeezing in the dB unit (panels a, c, and e) of the mechanical mode and the Bogoliubov mode, and the BEC-membrane entanglement (panels b, d, and f) vs normalized effective detuning $\delta_{c}/\kappa$ for three different values of $\xi_{2}/\xi_{1}\leq 0$. The lines indicated by $q (qs), Q (Qs)$, and $E_{N} (E_{Ns})$ correspond, respectively, to the quadrature squeezing of the mechanical mode, the Bogoliubov mode, and the BEC-membrane entanglement in the absence of the squeezed vacuum injection (presence of the squeezed vacuum injection with $N_{s}=10, \phi=\pi$) . In panels (a), (c), and (e) the thick horizontal line corresponds to the 3 dB limit. The other parameters are the same as those in Fig.~(\ref{fig:fig2}).
 	}
 	\label{fig:fig4}
 \end{figure}

\begin{figure}
	\centering
	\includegraphics[width=1.65in]{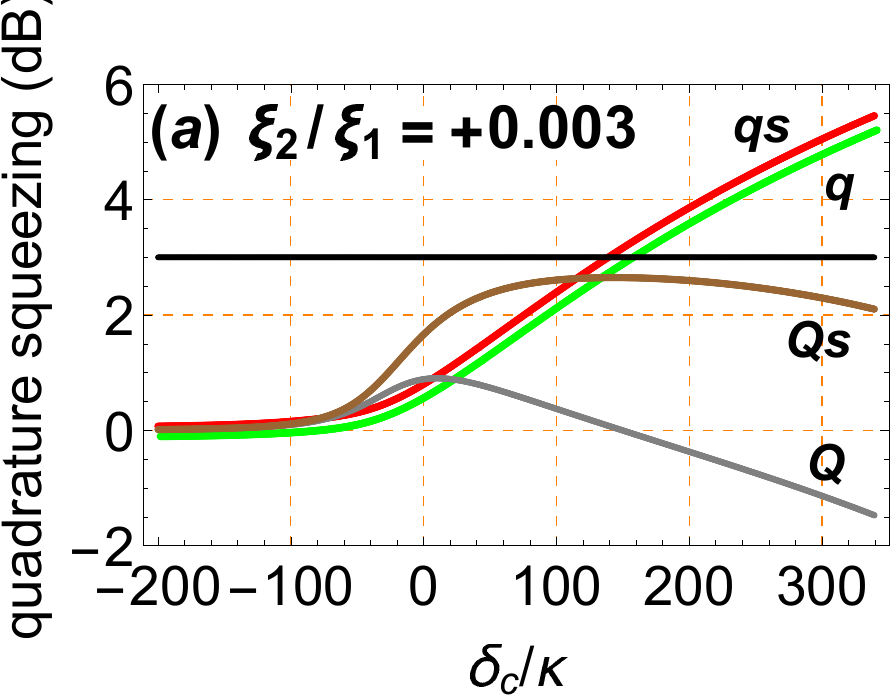}
	\includegraphics[width=1.65in]{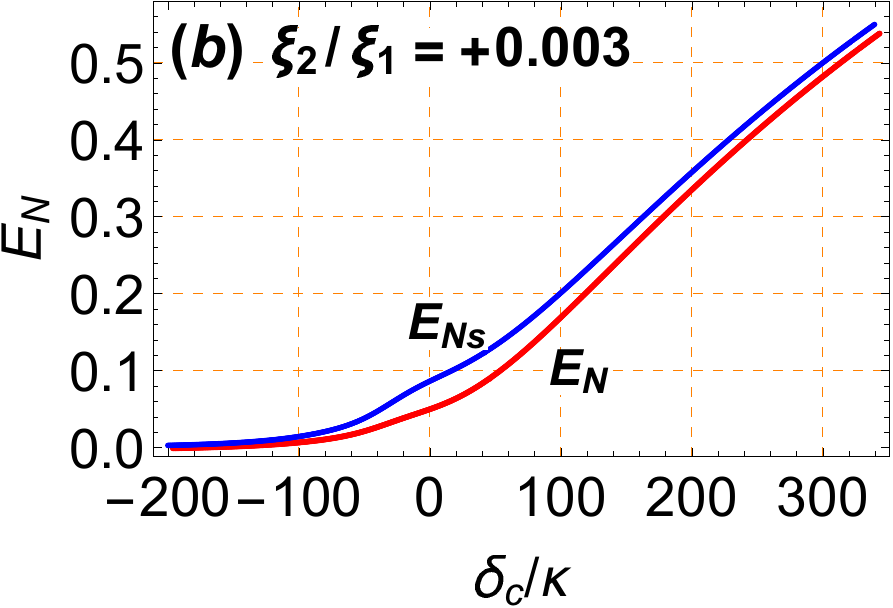}
	\includegraphics[width=1.65in]{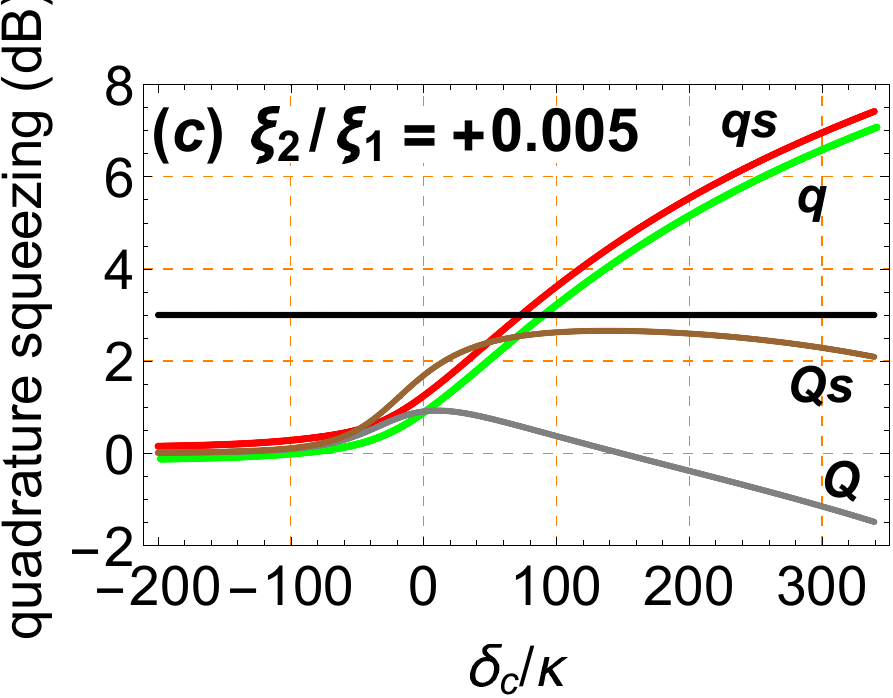}
	\includegraphics[width=1.65in]{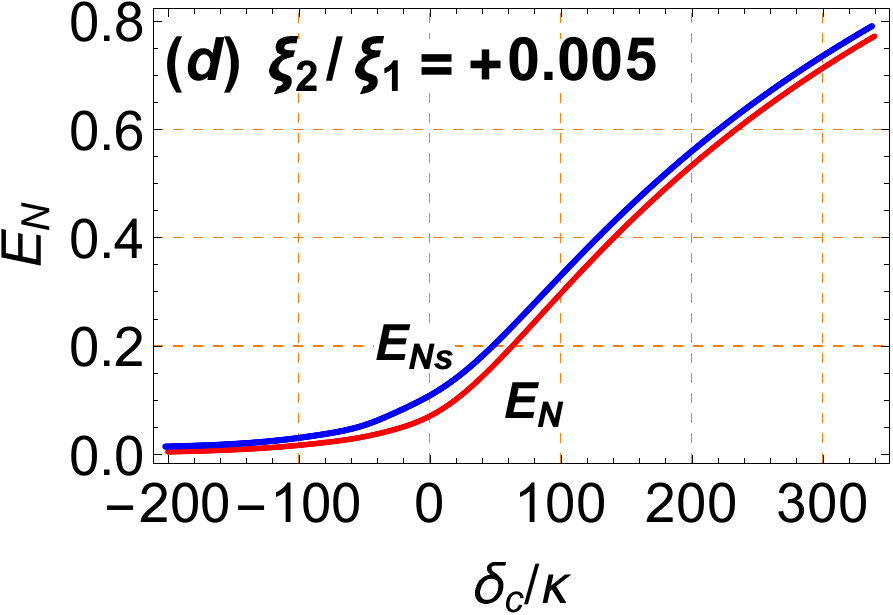}
	\includegraphics[width=1.65in]{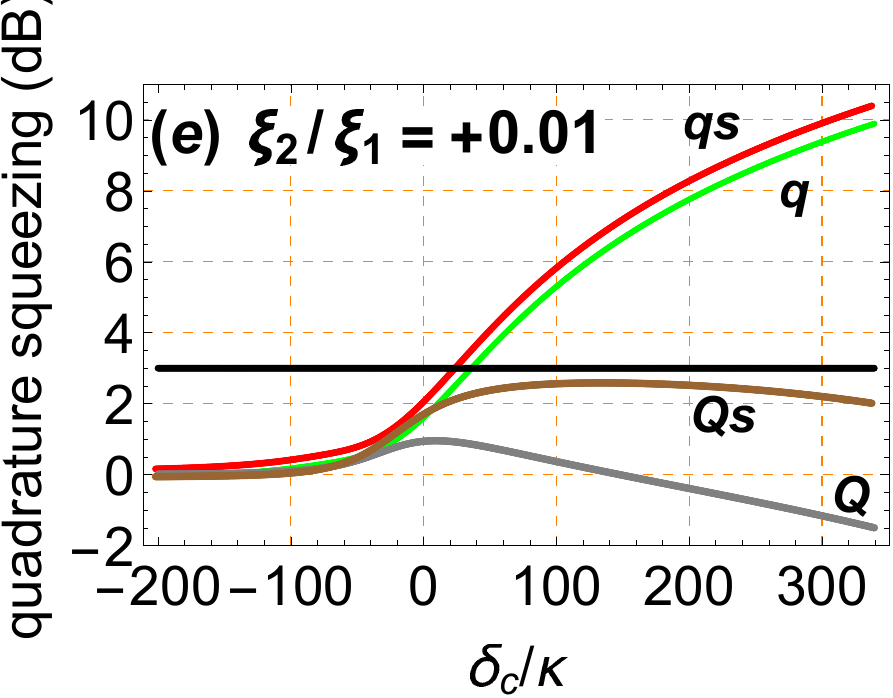}
	\includegraphics[width=1.65in]{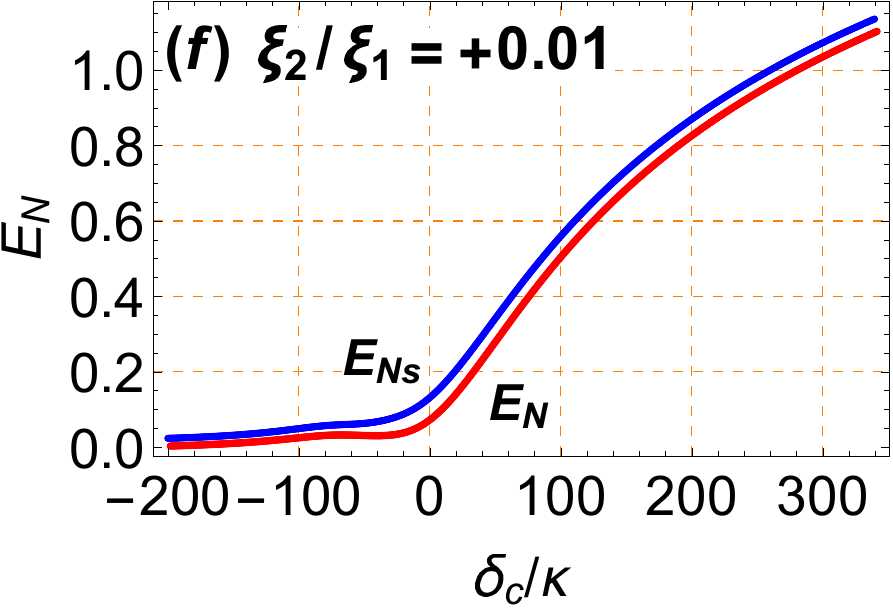}
	\caption{
		(Color online) The quadrature squeezing in the dB unit (panels a, c, and e) of the mechanical mode and the Bogoliubov mode, and the BEC-membrane entanglement (panels b, d, and f) vs normalized effective detuning $\delta_{c}/\kappa$ for three different values of $\xi_{2}/\xi_{1}> 0$. The lines indicated by $q (qs), Q (Qs)$, and $E_{N} (E_{Ns})$ correspond, respectively, to the quadrature squeezing of the mechanical mode, the Bogoliubov mode, and the BEC-membrane entanglement in the absence of the squeezed vacuum injection (presence of the squeezed vacuum injection with $N_{s}=10, \phi=\pi$) . In panels (a), (c), and (e) the thick horizontal line corresponds to the 3 dB limit. The other parameters are the same as those in Fig.~(\ref{fig:fig2}).
	}
	\label{fig:fig5}
\end{figure}

In Figs.~(\ref{fig:fig4}) and (\ref{fig:fig5}) we have shown the squeezing degree of the \textit{q}-(\textit{Q}-)quadrature of the mechanical (Bogoliubov) mode in dB as well as the BEC-membrane entanglement versus the normalized effective detuning $\delta_{c}/\kappa$ for $\xi_{2}=0$ [Figs.~\ref{fig:fig4}(a) and \ref{fig:fig4}(b)] and for two different negative values of the QOC, i.e., $\xi_{2}/\xi_{1}< 0$ [Figs.~\ref{fig:fig4}(c)-\ref{fig:fig4}(f)] and also for three different positive values of the QOC, i.e., $\xi_{2}/\xi_{1}> 0$ [Figs.~\ref{fig:fig5}(a)-\ref{fig:fig5}(f)] when the temperature of the system is $T=0.1\mu$K, the \textit{s}-wave scattering frequency of the atom-atom interaction is $\omega_{sw}=0.5\omega_{R}$ and the cavity is pumped at rate $\eta=100\kappa$. Also, it has been assumed that the parameters of the squeezed vacuum field have been fixed at $N_{s}=10, \phi=\pi$,

In those panels of Figs.~(\ref{fig:fig4}) and (\ref{fig:fig5}) showing the degrees of quadrature squeezing, the lines indicated by $Qs$ and those indicated by $Q$ correspond, respectively, to the \textit{Q}-quadrature squeezing of the Bogoliubov mode in the presence and absence of  the squeezed vacuum injection; the lines indicated by $qs$ and those indicated by $q$ correspond, respectively, to the \textit{q}-quadrature squeezing of the mechanical mode in the presence and absence of the squeezed vacuum injection. In addition, the thick horizontal lines denote the 3 dB limit. On the other hand, in those panels showing the BEC-membrane entanglement, the lines indicated by $E_{Ns}$ and those indicated by $E_{N}$ correspond, respectively, to the BEC-membrane entanglement in the presence and absence of the squeezed vacuum injection.

All the results obtained in Figs.~(\ref{fig:fig4}) and (\ref{fig:fig5}) have been calculated along the stable branch 1 of Fig.~(\ref{fig:fig3}). Based on our numerical calculations the phase of the vacuum injection field has the optimum value of $\phi=\pi$ which leads to the maximum BEC-membrane entanglement and also the maximum noise reduction in the mechanical and the Bogoliubov modes for each value of the effective detuning $\delta_{c}$. Besides, no more enhancement of squeezing and entanglement is achievable for $N_{s}>10$. That is why we have fixed $\phi=\pi$ and $N_{s}=10$ in Figs.~(\ref{fig:fig4}) and (\ref{fig:fig5}).

 As is seen from Figs.~\ref{fig:fig4}(a) and \ref{fig:fig4}(b), in the absence of the QOC and without squeezed vacuum injection no squeezing is observed in the mechanical mode and there is a very weak stationary entanglement between the BEC and the membrane ($E_{N}<0.04$). However, by injection of the squeezed vacuum field the mechanical mode is squeezed a little bit and the stationary entanglement between the BEC and the membrane increases to more than $E_{Ns}=0.1$. In addition, the squeezing degree of the Bogoliubov mode is increased over 2 dB in a wide range of the effective detuning [the line indicated by $Qs$ in Fig.~\ref{fig:fig4}(a)].
 
 On the other hand, in the presence of the QOC with negative sign [Figs.~\ref{fig:fig4}(c)-\ref{fig:fig4}(f)] the BEC-membrane entanglement in the absence and presence of the squeezed vacuum injection increases to more than $E_{N}=0.1$ and $E_{Ns}=0.4$ , respectively. Besides, by injection of the squeezed vacuum field one can increase the squeezing degree of the mechanical mode up to 2 dB and less than 3 dB, respectively, for $\xi_{2}/\xi_{1}=-0.003$ and $\xi_{2}/\xi_{1}=-0.005$. For more negative values of the QOC which have not shown here, the squeezing degree of the mechanical mode gets slightly larger than 3 dB only in the presence of the squeezed vacuum injection.
 
 These results reveal that for $\xi_{2}/\xi_{1}\leq 0$, i.e., in the absence of the QOC or in its presence with negative sign, by injection of the squeezed vacuum field one can enhance the BEC-membrane entanglement and also increase the squeezing degree of the mechanical and the atomic modes. In addition, for negative values of the QOC by increasing the absolute value of the QOC, i.e., $|\xi_{2}|$, both the BEC-membrane entanglement and the squeezing degree of the mechanical mode are increased. Nevertheless, beating the 3 dB limit is not possible for zero or negative values of QOC in the absence of the squeezed vacuum injection.

On the other hand, for $\xi_{2}/\xi_{1}=+0.003$ [Fig.~\ref{fig:fig5}(a)-\ref{fig:fig5}(b)] the squeezing degree of the Bogoliubov mode of the BEC in the presence of the squeezed vacuum injection can be reached to more than 2 dB for a wide range of the effective detuning while that of the mechanical mode of the membrane can be increased to more than 5 dB irrespective of the presence or the absence of the squeezed vacuum injection. Here, the noise reduction in the mechanical and the Bogoliubov modes lead to a very strong stationary entanglement between the BEC and the membrane which reaches to more than $E_{N}=0.5$ in the dispersive regime where the effective cavity detuning is much larger than the damping rate of the cavity  [$\delta_{c}>300\kappa$ before the system becomes unstable as is shown in Fig.~\ref{fig:fig5}(b)]. For lager positive values of the QOC [Figs.~\ref{fig:fig5}(c)-\ref{fig:fig5}(f)] the squeezing degree of the mechanical mode is increased to larger than 7 dB [Fig.~\ref{fig:fig5}(c)] and 10 dB [Fig.~\ref{fig:fig5}(e)] for  $\xi_{2}/\xi_{1}=+0.005$, and  $\xi_{2}/\xi_{1}=+0.01$, respectively, while that of the Bogoliubov mode is preserved at less than 3 dB. Furthermore, for these two values of $\xi_{2}/\xi_{1}$ the BEC-membrane entanglement is increased, respectively, to more than $E_N=0.7$ and $E_N=1$ even without the necessity of the squeezed vacuum injection

In short, the QOC can always enhances the stationary entanglement between the mechanical mode of the membrane and the Bogoliubov mode of the BEC irrespective of its sign. However, for positive values of QOC the BEC-membrane entanglement can be strengthened very intensively even without the necessity of squeezed vacuum injection. More interestingly, for the positive sign of the QOC the 3 dB limit of squeezing can be beaten for the \textit{q}-quadrature of the mechanical mode so that it may be squeezed to very high degrees even in the absence of squeezed vacuum injection for $\delta_{c}\gg\kappa$. On the other hand, the role of the squeezed vacuum injection is important in the enhancement of the \textit{Q}-quadrature squeezing of the Bogoliubov mode. Besides, the QOC with negative sign leads to the enhancement of both the BEC-membrane entanglement and the \textit{q}-quadrature squeezing degree of the mechanical mode. Nevertheless, beating the 3 dB limit is not possible for negative sign of QOC unless the cavity is injected with the squeezed vacuum field.

\section{Conclusions \label{Sec.VII}}
In conclusion, we have theoretically investigated a membrane-in-the-middle optomechanical cavity consisting of a cigar-shaped BEC of two-level atoms where the mechanical mode of the membrane  interacts both linearly and quadratically with the radiation pressure of the cavity. The cavity is driven through the left mirror by an external laser and also is injected by a squeezed vacuum field with a central frequency which is assumed to be at resonant with the cavity mode. If the laser pump is far detuned from the atomic resonance the atomic wave function can be described by a single-mode scalar quantum field in the Bogoliubov approximation which is coupled linearly to the radiation pressure of the cavity.

We have shown that the QOC not only affects the dynamical behavior of the mean fields but also changes the optical bistability behavior to tristability behavior and also modifies the steady-state quantum fluctuations of the system. It has been shown that for negative values of the QOC, the transition threshold from bistability to tristability can be controlled by the absolute value of the QOC. Furthermore, the QOC modifies the effective oscillation frequencies of the mechanical and the Bogoliubov modes and also their relaxation times.

On the other hand, the quantum fluctuations of the system can be manipulated by both the QOC and the squeezed vacuum injection. It has been shown that by suitable combination of the QOC and the vacuum field injection one can increase the noise reduction in the mechanical and the Bogoliubov modes to very high extent and produce a very strong stationary-state entanglement between them. More interestingly, for positive values of the QOC one can achieve a very high degree of squeezing in the mechanical mode and a strong entanglement between the mechanical and atomic modes without the necessity of the squeezed light injection.

\section*{Acknowledgment}
A. D wishes to thank the Laser and Plasma Research Institute of Shahid Beheshti University for its support.

\bibliographystyle{apsrev4-1}

\end{document}